\documentclass[final,3p,times,twocolumn,authoryear]{elsarticle}

\usepackage{stfloats}
\usepackage{amssymb}
\usepackage{amsmath}
\usepackage{graphicx}
\usepackage{caption}
\usepackage{booktabs}
\usepackage{wrapfig}
\usepackage{subfigure}
\newcommand{\tabincell}[2]{\begin{tabular}{@{}#1@{}}#2\end{tabular}}
\usepackage[colorlinks,
linkcolor=red,
anchorcolor=block,
citecolor=green]{hyperref}
\usepackage{natbib}
\setcitestyle{authoryear,round}

\journal{Nuclear Physics B}

\begin{document}
\begin{frontmatter}
		
		\title{RC-chain: Reputation-based Crowdsourcing Blockchain for \\Vehicular Networks}

\author[1,2]{Lijun Sun\corref{cor1}}
\cortext[cor1]{Corresponding author}
\ead{lijunsun@qust.edu.cn}
\author[1]{Qian Yang}
\ead{yangqian514v@163.com}
\author[3]{Xiao Chen}
\ead{xiao.chen@ed.ac.uk}
\author[2]{Zhenxiang Chen}
\ead{czx@ujn.edu.cn}
\address[1]{College of Information Science and Technology, Qingdao University of Science \& Technology, Qingdao 266061, China}
\address[2]{Shandong Provincial Key Laboratory of Network Based Intelligent Computing, University of Jinan 250022, China}
\address[3]{School of Informatics, University of Edinburgh, Edinburgh, EH8 9AB, UK}

\begin{abstract}

\par As the commercial use of 5G technologies has grown more prevalent, smart vehicles have become an efficient platform for delivering a wide array of services directly to customers. The vehicular crowdsourcing service (VCS), for example, can provide immediate and timely feedback to the user regarding real-time transportation information. However, different sources can generate spurious information towards a specific service request in the pursuit of profit. Distinguishing trusted information from numerous sources is the key to a reliable VCS platform. This paper proposes a solution to this problem called “RC-chain”, a reputation-based crowdsourcing framework built on a blockchain platform (Hyperledger Fabric). We first establish the blockchain-based platform to support the management of crowdsourcing trading and user-reputation evaluating activities. A reputation model, the Trust Propagation \& Feedback Similarity (TPFS), then calculates the reputation values of participants and reveals any malicious behavior accordingly. Finally, queueing theory is used to evaluate the blockchain-based platform and optimize the system performance. The proposed framework was deployed on the IBM Hyperledger Fabric platform to observe its real-world running time, effectiveness, and overall performance.

\end{abstract}

		
\end{frontmatter}

\section{Introduction}
\label{sec1}
Enhanced wireless communication (e.g., 5G) technologies expand the scope of vehicular network services, such as the vehicular crowdsourcing service (VCS), which leverages smart device-equipped vehicles to support the sharing and trading of information among vehicles \citep{wu2013location}. In the VCS scenario, behaving as either service consumers or providers, vehicle-nodes cooperatively collect and share data of common interest (e.g., road conditions of a given destination, parking space availability). A specific request may be considered a payable mission that is assigned to a group of service providers via crowdsourcing platform \citep{howe2006rise}. Once the mission is completed by the service providers, an optimal solution is selected by the service consumer via information sharing: the selected service provider is compensated for completing the trade. Service providers can also proactively share traffic jam warnings, or recommend restaurants along a certain route. The VCS when functioning optimally, gives users a safe, convenient and comfortable driving experience.

In addition to these advantages, the VCS is markedly advantaged by security threats. Researchers have introduced blockchain technology into vehicular networks in effort to exploit decentralization, immutability, transparency, and traceability characteristics \citep{huang2020securing}. However, the entrance of a host of vehicle-nodes to the permissionless public blockchain exacerbates security and privacy risks, not to mention the high costs and high latency caused by the mining process of miners. By contrast, the consortium blockchain has strict permissioned standards for vehicle-nodes, which to a certain extent guarantees the legality of disseminated information \citep{kang2018blockchain}. However, the consortium-based VCS will face challenges when encountering the following scenarios.

\begin{itemize}
	\item Scenario 1: All nodes that are permitted to join the consortium blockchain are legal, but this does not mean that their behavior is always regulated. Free-riding, spoofing, and repudiation behaviors, for example, may happen occasionally. What is the best approach to eliminating selfish or malicious behaviors of vehicles? In other words: how does one engineer a trusted platform that allows users to trade information without worrying about cheating or attacks? 
	\item Scenario 2: When a VCS mission is released, many vehicles make requests indicating whether or not they can provide services. How does one distinguish the most valuable information from the numerous information sources? 
	\item Scenario 3: The “reputation value” concept was introduced to solve the above problems as a basis for selecting the best service providers \citep{yang2018blockchain,kang2018blockchain}. The reputation value can be calculated through mutual ratings between vehicles based on the assumption that all vehicles provide honest ratings for others. However, the value loses its meaning when malicious ratings or false reporting occur. So, how can these untruthful rating behaviors be eliminated?
	\item The limited resources of traditional centralized crowdsourcing platform are also taxed as the quantity of vehicle-nodes increase, coupled with the rapid growth of dissemination information. Large-scale user groups pose challenges in building an efficient VCS platform.
\end{itemize}

This paper proposes a reputation-based crowdsourcing blockchain framework, the RC-chain, which was designed to address these problems. The RC-chain implements crowdsourcing management on a blockchain platform, the IBM Hyperledger Fabric \citep{fabric-whitepaper}, which is transparent, irreversible and decentralized. Each iteration of crowdsourced information-trading is recorded as a transaction of the blockchain based on a group of agreed-upon criterion specified in the smart contract (i.e., the chaincode in Fabric), which prevents adversarial user behaviors and guarantees a secure trading environment. A reputation model is also established with the RC-chain through smart contract to improve the crowdsourcing service quality. The reputation model is used to evaluate the trust-degree of VCS users, both service consumers and providers, based on their activities - including the quality of mission completion with service providers and the payment-operations of service consumers. This quantified reputation value can help consumers decide the potential providers and also can help providers choose their missions. The reputation value of the service vehicle is updated and recorded in the chain each time a crowdsourcing service is completed. When there is a problem with the reputation of a certain vehicle, it is marked with a “warning”.

We integrated our designed information-trading and reputation-evaluating components, to implement RC-chain atop the IBM Hyperledger Fabric, which combines the transaction process and a three-stage consensus. We used queueing theory to optimize configuration and performance of the Fabric-based RC-chain. Numerical analysis and practical experiments were conducted to test the feasibility, efficiency, and overall performance of the proposed framework. The primary contributions of this work can be summarized as follows.

\begin{itemize}
	\item A novel blockchain-based VCS platform is proposed for enhanced trust, the Hyperledger Fabric-based RC-chain. Each trading record is designed as a RC-chain transaction by specifying all trading rules in transparent smart contract.
	\item A novel reputation model is designed and aggregated with the RC-chain to improve crowdsourcing service quality. The model prevents adversarial or irresponsible behaviors in handling crowdsourced missions by degrading the reputation of any malicious service consumers or providers. Additionally, a Trust Propagation \& Feedback Similarity (TPFS) model identifies malicious service providers by analyzing their mission-based reputation figures: this helps consumers find an optimal service provider to receive high-quality service.
	\item Queueing theory is applied to generate an efficient configuration scheme that enhances RC-chain platform performance. We conducted a series of experiments on key performance indicators such as transaction confirmation time and transaction throughput to further assess the proposed framework’s effectiveness.
\end{itemize}

To the best of our knowledge, the RC-chain is the first solution to combine blockchain-based crowdsourcing management with reputation evaluation and queueing-based optimization approaches. Our analysis, as discussed in detail below, indicates that the framework allows for secure management and major performance improvement in VCS scenarios.

Below, Section~\ref{sec2} discusses previous work related to this study. Section~\ref{sec3} describes the VCS scenarios, roles and deployment relevant to RC-chain. Section~\ref{sec4} introduces the transaction process and consensus mechanism. An optimized reputation management model for calculating the reputation value of participants is given in Section~\ref{sec5}. We modeled the RC-chain using queueing theory and evaluated its performance as discussed in Section~\ref{sec6}. The results of the experiment and our interpretation of them are presented in Section~\ref{sec7}. Section ~\ref{sec8} gives concluding remarks.

\section{Related work}
 \label{sec2}
 Trust is an important research topic in the vehicular network field. To solve the challenge of the trust problem, many studies have been carried out. Existing trust models can be divided into three categories based on their respective evaluation objects: entity-centric trust models that focus on vehicle credibility, data-centric trust models that focus on received messages' credibility, and combined trust models for evaluating both entities and data \citep{lu2018privacy}. Existing trust management methods are mainly centralized \citep{chen2017cloud,tian2019vcash}, or distributed \citep{haddadou2014job,zhang2016distributed}.

Many previous researchers have explored trust management schemes for Internet of Vehicles (IoV). Chen and Wang \citep{chen2017cloud}, for example, used cloud computing technology to build a three-tier trust model to ensure the reliability and security of reputation values. Vcash was introduced as a reputation framework for identifying denial of traffic service \citep{tian2019vcash}; it relies on a central server, however, resulting in centralization problems. Vcash also assumes that the Road Side Units (RSUs) are trustable, which actually may provide false messages owing to being attacked \citep{zhang2011survey}. \citet{raya2006securing} conducted a comprehensive analysis of attack problems. Cloud servers have high communication costs and usually cannot satisfy delay requirements, so it is difficult to guarantee the quality of service \citep{greenberg2008cost}.

 An alternative approach is the distributed aggregate privacy-preserving authentication (DAPPA) design \citep{zhang2016distributed}. Its necessary generation of one-time private key and aggregate signatures results in latency issues and affects system performance \citep{lu2018survey}. \citet{haddadou2014job} developed the distributed trust model DTM2 to detect and eliminate malicious nodes, but could not ensure the safety of the reputation system itself. Other researchers have utilized the behavior-based trust management mechanism \citep{marmol2012trip,wex2008trust} to establish entity-centric models, which guarantees the accuracy of the reputation value, but does not secure the reliability of updated reputation values during storage. At present, there are several major challenges for electronic data storage. Unilateral deposit schemes are costly, allow data to be easily lost or tampered with, and are difficult to trace.
\begin{table*}[ht]
	\label{tab1}
	\centering

	\begin{tabular*}{15cm}{lcccc}  
		\multicolumn{3}{l}{\small{\textbf{Table 1}}}\\
		\multicolumn{3}{l}{\small{Comparison of the RC-chain with the existing schemes}}\\
		\hline  
		Reference & Permission & Distribution & Incorrect Recommendations & Feedback Similarity  \\  
		\hline  
		\citep{kang2018blockchain} & \checkmark &\checkmark&$ \times $&$ \times $ \\  
		\citep{yang2018blockchain} & $ \times $ &\checkmark&$ \times $&$ \times $ \\ 
		\citep{jabbar2020blockchain} & $ \times $ &\checkmark&$ \times $&$ \times $ \\ 
		\citep{tian2019vcash} & $ \times $ &$ \times $&$ \times $&$ \times $ \\
		\citep{xiao2020edge} & \checkmark &\checkmark&$ \times $&$ \times $ \\
		RC-chain(This paper) & \checkmark &\checkmark&\checkmark&\checkmark \\
		\hline  
	\end{tabular*}  
\end{table*} 

Blockchain technology has attracted extensive attention from researchers for its qualities of decentralization, autonomy, traceability and non-tampering modification. It has emerged a viable solution to store and update reputation values. \citet{yang2018blockchain} established a decentralized trust management system for vehicle networks based on blockchain technology. It adopts a consensus algorithm that combines Proof of Work (PoW) and Proof of Stake (PoS), where RSUs with larger stakes are more likely to be elected as a block producer. However, the miner hash value calculation process requires excessive computing power; few individuals are paid while most are simply wasting their resources \citep{wang2018survey}. This is not desirable for vehicles with limited resources. Additionally, the consensus mechanism is collaboratively implemented by participants with conflicting interests. In a city or local area, RSUs is usually deployed and maintained by one or two network service providers, which creates potential security risks (e.g., 51\% attack) \citep{chen2020}. The PoS scheme is also susceptible to “rich getting richer" problems \citep{zheng2017overview}.

\citet{li2018creditcoin} proposed an incentive network based on privacy protection. CreditCoin, which uses a Byzantine fault-tolerance (BFT) consensus algorithm to encourage vehicles to publish messages. \citet{awais2019secure} established security and trust in a decentralized vehicular environment without RSUs, by focusing on blockchain-based message distribution through short-range communication protocol. Their technique was also shown to ensure fast, secure transmission and accurate recording of the data. \citet{jabbar2020blockchain} built the Decentralized IoT Solution for Vehicles communication (DISV) by adapting blockchain technology for real-time application (RTA).

The above approaches involve unresolved issue in terms of public blockchain schemes \citep{yang2018blockchain,li2018creditcoin,singh2017blockchain}. Insensitivity to delay requirements is a problem, for example, because the public chain has no access permission requirements - any node can participate in the consensus process \citep{zheng2017overview}, resulting in block generation too slow for certain real-time reply services in vehicular networks. Privacy issues also merit concern. Data transmitted and stored publicly, does not comply with the requirements involving commercial secrets \citep{ali2018applications}. General vulnerability to attacks is a noteworthy problem as well - driven by self-interest, malicious nodes can easily join the public blockchain network without permission. A consortium blockchain may be able to resolve these issues. \citet{kang2017enabling}, for example, built a peer-to-peer (P2P) power trading system that requires no third party based on consortium blockchain technology.

Additionally, \citet{zou2018proof} proposed the consensus protocol Proof-of-Trust (PoT) using a blockchain for crowdsourcing service. PoT consensus is a hybrid blockchain architecture that integrates a consortium blockchain in an open public service network. They focused on improving the consensus protocol without designing new transaction processes or crowdsourcing scenarios, such as IoV. \citet{xiao2020edge} proposed a network computing framework Quick Fake News Detection (QcFND) to quickly detect fake news in IoV. QcFND used a data-centric, instead of entity-centric, trust management scheme based on Bayesian networks.

\citet{kang2018blockchain} used edge computing and consortium blockchain for secure P2P data sharing. A reputation-based data sharing scheme with three-weight subjective logic (TWSL) was developed to select reliable data sources. However, using a PoW consensus scheme can waste resources and create “double spending” problems. The TWSL can be further improved in identifying untruthful ratings, as was one of our main goals in conducting this study. Relatively few blockchain researchers have explored theoretical modeling of the system, though, \citet{memon2019simulation} built a queueing theory-based model to evaluate ideal transactions statistics in Bitcoin and Ethereum. We summarize the most related works in \hyperref[tab1]{Table 1}.

Similarly, we applied the consortium blockchain to a vehicular network crowdsourcing service in this study. We utilized the fog computing paradigm to improve quality of service (QoS), and developed the RC-chain framework to meet the trust challenge. A detailed description of our framework is given below.

\begin{figure*}[ht]
	\centering
	\includegraphics[scale=0.4]{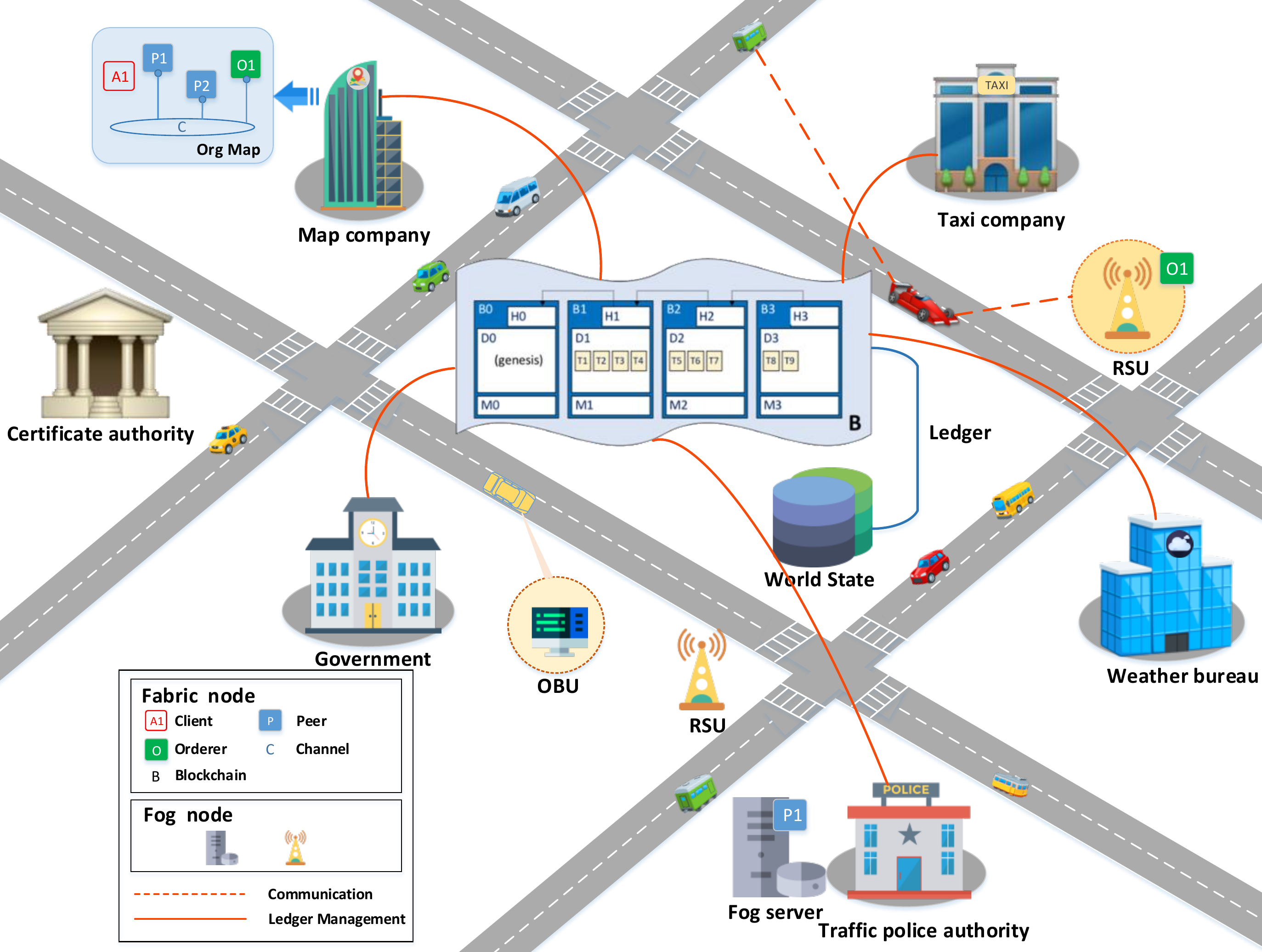}
	\caption{Regional consortium of VCS with RC-chain.}
	\label{fig1}
\end{figure*}
\section{Framework description}
\label{sec3}
\par We built the proposed RC-chain based on consortium blockchain and fog computing technology to support a trusted environment and high QoS for VCS. We added “organization” and “crowdsourcing” to the traditional vehicular network to fully utilize the infrastructure and resources.
\subsection{Scenario overview}
 \label{sec3.1}
\par The scenario consists of the following entities: Certificate Authority (CA), organizations, blockchain, vehicles, RSUs, and fog servers (\hyperref[fig1]{Fig. 1}). Various organizations can form a regional consortium based on RC-chain. We divide the crowdsourcing services into two types here, the question and answer (Q\&A) service and the data-sharing service. 

 The Q\&A service involves a provider answering to questions proposed by the service consumer. For instance, consider a map company that needs traffic condition information for a certain road at a specific time, a weather bureau that needs to collect the precise air quality status of a park, or a traffic police authority that needs to know in real-time where traffic accidents have occurred. As potential service consumers, the vehicles within these organizations ask questions, then nearby vehicles including taxis, buses, and passenger cars can compete as service providers to answer the questions.
 
 The data-sharing service refers to vehicles actively sharing data among various organizations. For example, a taxi service provider publishes collected data which can be utilized by other organizations (e.g., a traffic police authority) to analyze and plan holiday route changes. The weather bureau shares weather predictions which can be shared by the taxi company with their passengers. More details about this type of transaction process are introduced in Section~\ref{sec4.1}.

RC-chain has a three-layer network architecture composed of a user stratum, fog stratum, and cloud stratum. The user stratum includes on-vehicle sensors and terminal devices which upload the collected data to the fog nodes, that is, the RSUs distributed along the road and organization-owned fog servers. As the number of transactions continues to increase, the fog nodes can upload early accounts to the cloud stratum after a certain period. Organizations can rent cloud services from suppliers based on their own needs to conduct big data analysis and large-capacity storage as necessary.

\subsection{Roles of assignment}
 \label{sec3.2}
\par The roles and functionalities of entities are as follows.

\textbf{Certificate Authority}: The CA issues certificates and often held by the audit and management department. Unlike the traditional P2P trading of public blockchains, nodes without identity certificates are not permitted to participate in the transaction. Each user obtains a legal identity, then is given an initial reputation value that is constantly updated based on its service behavior (Section~\ref{sec5}). The vehicle’s subsequent behavior after joining the consortium is reflected through the transaction records in the ledger \citep{lu2018privacy}.

\textbf{Organizations}: An organization is a collection of multiple members who generally share the same root certificate. An organization can be as large as a multi-national corporation or as small as an individual, e.g., traffic police authority, taxi company, map company, or weather bureau as shown in \hyperref[fig1]{Fig. 1}. A collection of organizations with common interest, joining to a network only by authorization, form a consortium. Nodes in the organization, such as vehicles, are the transaction objects.

 \begin{figure*}[ht]
	\centering
	\includegraphics[scale=0.2]{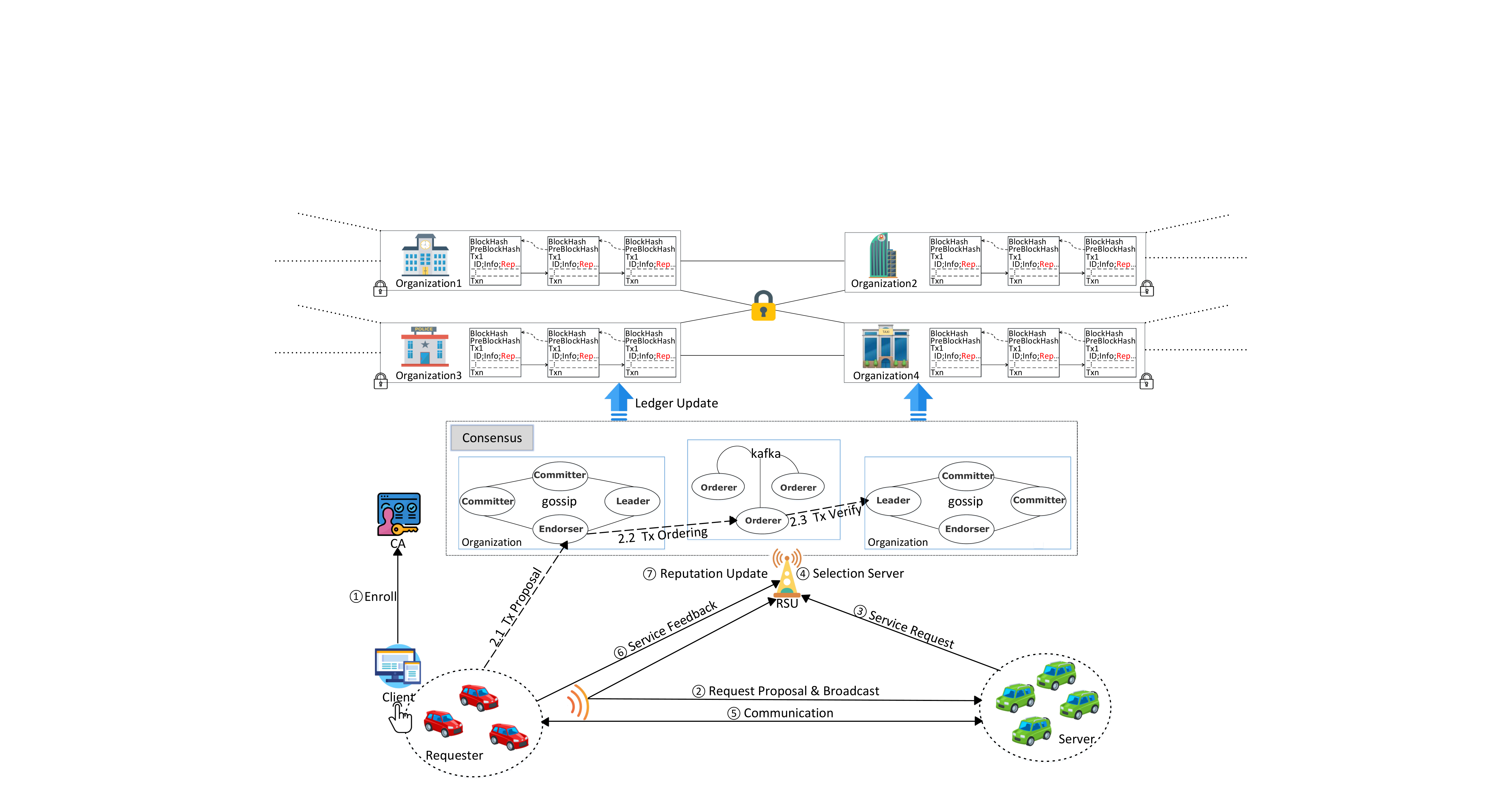}
	\caption{Transaction process.}
	\label{fig2}
\end{figure*}
\textbf{Blockchain}: The Hyperledger Fabric platform of the consortium blockchain is an important component of RC-chain, which consists of three main components: peers with the ledger, orderers, and clients. Peers may be endorsing peers, committing peers, or leading peers. Each organization has its peers with copies of the ledger of their consortium network. The ledger is composed of the transactions logs of the blockchain structure and the world state, the database of the ledger, which describes the state of the ledger at a given point in time. The transactions logs record all transactions that produce the current value of the world state regardless of legality. Orderers are responsible for ordering transactions and packaging them into blocks. The client is operated by users. Every organization may contain a different number of peers, orderers, and clients. As shown in \hyperref[fig1]{Fig. 1}, the map company with two peers and one orderer initiates transactions through client A1. The peers and orderers work together to complete the consensus process; endorsing peers, orderers, and committing peers can all run independently for enhanced efficiency (Section~\ref{sec4.2}).

\textbf{Vehicles}: Each vehicle belongs to one organization as a user capable of playing different roles (requester, server, idler) according to their own VCS needs. The client is deployed on vehicles. Vehicles are equipped with advanced communication devices such as On-Board Units (OBUs) which can collect data through sensors, execute simple calculation and storage operations, and communicate with nearby RSUs \citep{lasla2018efficient}.

\textbf{RSUs}: A certain number of RSUs are distributed in each area. RSUs act fog nodes that collect transaction requests and update reputation values. Orderers are deployed on RSUs to order services. The RSUs executes these processes according to agreed-upon rules specified in the smart contract, and upload the process to the ledger to be supervised by other nodes.

\textbf{Fog servers}: By leveraging the rich computing and storage resources of fog servers close to the edge of the network, peers deployed on fog servers within various organizations can effectively minimize latency. Transaction information is stored in the distributed ledger of the fog server for the purposes of consistency and transparency. 

\section{RC-chain design}
 \label{sec4}
\par In this section, the specific transaction process integrated on the RC-chain is elaborated, and it works with the three-stage consensus to achieve platform-level and service-level security.
\subsection{Transaction process}
 \label{sec4.1}

All transactions are performed via chaincode. A vehicle broadcasts the request for Q\&A service as shown in \hyperref[fig2]{Fig. 2}. The request proposal and following main process need verifications of three-stage consensus to proceed. Then they are recorded in the ledger as evidence of transactions. Any vehicle willing to answer the question can send a service request to the nearest RSU, which selects the server according to TPFS model. After the service is finished, the requester evaluates the service, then the reputation value of the corresponding server is updated. The data-sharing service only requires the server to upload the shared data index. The requester selects the data of interest according to the index, then establishes communication with the server to obtain integral data, other processes are similar to the Q\&A service.

\textbf{Step 1}: All participants joining in the consortium should be authenticated in the CA by binding their registration information (e.g., vehicle license plate). CA issues public keys, private keys, transaction certificates, and identity certificates to legal entities via an elliptic curve digital signature algorithm and asymmetric cryptography. The initial reputation value of entities is generally 0.5.

\textbf{Step 2}: The requester broadcasts the request to the other vehicles and nearest RSU (the “leading” RSU), which communicates with other (“following”) RSUs in the same area to keep in sync. The following RSU passes the request to vehicles in its vicinity. The requester sends the request transaction proposal to endorsing peers in the fog servers of organizations to execute consensus. Endorsing peers are chosen by the endorsement policy in chaincode, which defines organizations that need to endorse the proposal to be accepted by the ledger. This is the literal definition of “consensus” – every organization involved must endorse a proposed ledger before it can be accepted by any peer of any organization. Vehicles must reach consensus before a transaction proposal can be completed. After the proposal passes the three stages of endorsement (Step 2.1), ordering (Step 2.2), and commitment (Step 2.3), it is considered to be a legal transaction and the ledger can be updated accordingly (Section~\ref{sec4.2}).

\textbf{Step 3}: The service-intentioned vehicle submits a service-request to the nearest RSU.

\textbf{Step 4}: The leading RSU and following RSUs select one service candidate based on the TPFS model (Section~\ref{sec5}). The following RSUs send the service candidates to the leading RSU, who selects the final server at random. The final server sends a service proposal with the RSU’s signature to the ledger.

\textbf{Step 5}: The requester establishes communication with the server, then the server sends the service-process proposal to the ledger as a proof of transaction.

\textbf{Step 6}: After the service is completed, the requester gives the leading RSU feedback (e.g., rating, timeliness report) about the server.

\textbf{Step 7}: The RSUs update the reputation value of the servers according to the TPFS model and send a reputation-update proposal to the ledger. In this way, the reputation value can be traced through the transaction records of the block.

\subsection{Three-stage consensus}
 \label{sec4.2}
 In the blockchain, “consensus” is an agreement on the order and status of multiple transactions generated within a certain period. The transaction is packaged into blocks according to certain rules to ensure that the ledger owned by the distributed nodes retains a stable state. It is possible that not all nodes run in an ideal state however, so a current consensus algorithm of the blockchain system is very challenging to establish. For example, this probabilistic consensus algorithm PoW causes not only a waste of resources but also “double spending” problems possibly. The PoS is susceptible to collusion attacks by malicious nodes. The Practical Byzantine Fault Tolerance (PBFT) cannot prevent sybil attacks, that is, where malicious users exploit multiple IDs to participate in erroneous consensus behavior. The pursuit of security can cause problems such as long delays and poor scalability.

 Fortunately, the consensus of the Hyperledger Fabric is not susceptible to double spending. The three-stage consensus can effectively ensure security. It is the consensus of the whole process based on staged consensus, that is, the failure of any stage will lead to failure in the end. The separation between endorsement and ordering lends performance and scalability advantages, while the use of fog computing technology accelerates the processing time. 
 
\begin{figure}[ht]
 	\centering
 	\includegraphics[scale=0.3]{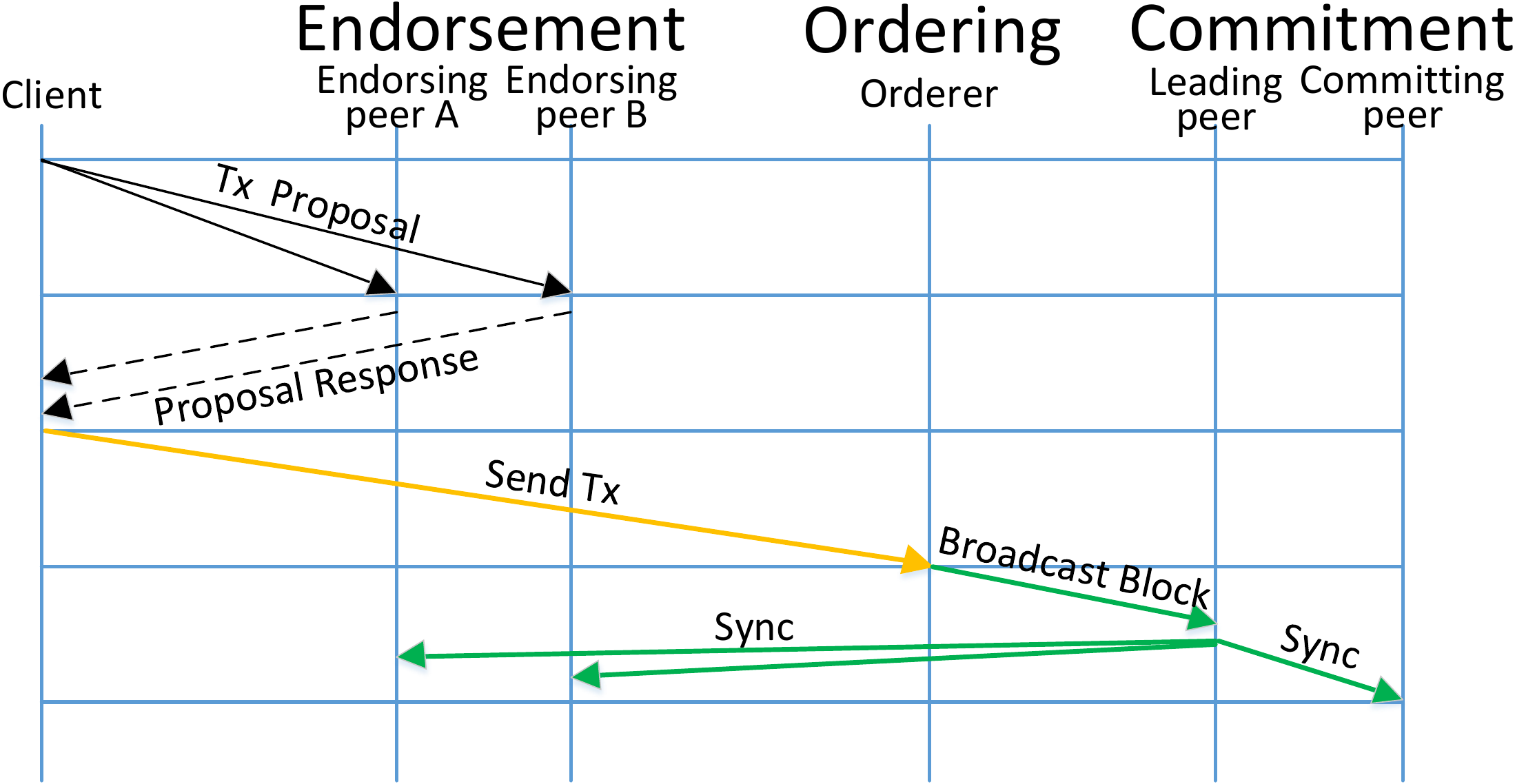}
 	\caption{Three-stage consensus.}
 	\label{fig3}
 \end{figure}

 The transaction proposal discussed in Section ~\ref{sec4.1} can only be written into the ledger after the endorsement, ordering, and commitment steps (2.1-2.3 in \hyperref[fig2]{Fig. 2}) of consensus are all complete. As shown in \hyperref[fig3]{Fig. 3}, there are two types of nodes participating in the consensus process. Peers are mainly used for endorsement and commitment and are divided into endorsing peers, leading peers, and committing peers categories while orderers are used for ordering.

 Endorsement stage: The vehicles exploit the client to send the transaction proposal to the designated endorsing peers on fog servers of related organizations. The endorsing peers comprise a small portion of the total peers and are dynamically designated by the endorsement policy. Compared to situations wherein all nodes participate in the endorsement process, this markedly improves efficiency. Upon receiving a transaction proposal, the endorsing peers simulate execution of the transaction process per the chaincode, and record results, but the status of the ledger is not yet updated, The endorsing peers of various organizations then return the transaction proposal response with signatures to the client.

 Ordering stage: After receiving sufficient endorsement, the client sends the endorsed transaction proposal to the orderers on RSUs. An orderer does not read the transaction content, but only orders the received transactions using the Kafka algorithm in real time, then packs them into blocks and sends them to the leading peer.

 Commitment stage: The most basic function of peers is commitment. The leading peer in each organization synchronizes the received block to other committing peers. The committing peers verify the validity of transactions in the block, including the transaction structure, signature integrity, whether it is repeated, and whether the read and write collection versions match. (This step may contain a double spending transaction, which is considered invalid.) After verification, the transaction is written to the updated ledger.

The three-stage consensus performs well in fault tolerance. In the endorsement stage, the clients learn the current transaction status through feedback from the endorsing peers. The Kafka algorithm in the ordering stage supports Crash Fault-Tolerance (CFT), the downtime of a few orderers does not affect the generation of blocks. If a committing peer fails to synchronize a block due to communication failures, etc., it can pull the block from other committing peers to maintain ledger consistency. If transactions (e.g., the update of reputation value) can not be completed due to the consensus failure or network breakdown, they are not updated in the ledger, but recorded in the transactions logs for audit.

 Due to the mobility of the vehicle, the client may have diconnection problem. In the endorsement stage, if client does not receive sufficient endorsement signatures, there are two possible cases:

 \textbf{Case 1}. The client did not send the transaction proposal to sufficient endorsing peers for endorsement.
 
 \textbf{Case 2}. Execution timeout of endorsement peers and no response. 
 
 For the Case 1 the client can propose the transaction again. Though Case 2 rarely occurs, we take it into account. The risk of disconnection can be reduced by setting in the endorsement policy. For instance, 1) increase the number of endorsing peers; 2) allow for passing the verification with fewer signatures; 3) select more backup nodes in client’s vicinity, etc. Even though they are all disconnected, repeat the submission.

The three-stage consensus process thus achieves BFT.
 
\section{Reputation model}
 \label{sec5}
 Our optimized reputation model is introduced in this section. Many researchers have established reputation value management models. For users who have joined the consortium, it is necessary to improve QoS in order to identify their possible selfish behavior, further help to choose higher-quality services. In addition, reputation can also be used as an incentive to encourage participants to take part in the interaction through reputation-based methods. Those users who contribute more reliable services will get more service opportunities and rewards. \citet{kang2018blockchain} took interaction frequency, event timeliness, trajectory similarity, and recommended opinions into account. Subjective logic is utilized to formulate the individual reputation evaluation based on occurring interactions, which is a framework for probabilistic information fusion operated on subjective beliefs about the world. Based on the TWSL \citep{kang2018blockchain}, we improved the recommended opinions factor as the so-called trust propagation (TP). Inspired by \citep{liu2016machine}, we applied feedback similarity (FS) for the reputation model. As mentioned in Section~\ref{sec4}, the TPFS model (which encompasses both TP and FS factors) is integrated into the chaincode to update the reputation values. We analyzed the TPFS model and compared it against the TWSL model as discussed in detail below.

\subsection{Trust propagation}
 \label{sec5.1}
 When the number of vehicles is very sparse in the vehicular network, most have not established interactions and thus it is difficult to find a suitable vehicle to send an inquiry request. The recommendations of other vehicles can be considered here, namely, TP. If $  i  $ has a reputation value to $  j  $ and $  j  $ has a reputation value to $ f $, then $  i  $ can calculate the reputation value to $ f $ through $  j  $. TWSL synthesizes the general opinions of all vehicles on $ f $, but does not consider any situation wherein vehicles with lower reputation values make incorrect recommendations. Here, we do take this into consideration to distinguish between reliable and unreliable opinions.

We set two thresholds $ {T}_l, {T_h} $ for the reputation value, and deem the recommendation of neighbors with reputation values lower than $ {T}_l $ to be unreliable while those of vehicles with recommendation values higher than $ {T}_h $ are reliable. We define the \textit{recommended confidence} of vehicle $ i $ to $ j $ as:

\begin{equation}
\label{equ1}
{{{C}}_{i \to j}}{{ = }}\left( \begin{array}{l}
	0,\qquad{{{R}}_{i \to j}} < {T_l}\\
	0.8,\quad{{{T}}_l} < {R_{i \to j}} < {T_h}\\
	1,\qquad{{{R}}_{i \to j}} > {T_h}
\end{array} \right.
\end{equation}
where $  {{R}}_{i \to j} $ is the \textit{reputation value} of neighbor vehicle $ i $ to vehicle $ j $.
\par The recommendation of vehicle $  j  $ can be divided into positive and negative opinions. For a positive opinion, the reputation value $ {R_{j \to f}} $ of neighbor vehicle $ j $ to vehicle $ f $ is greater than reputation threshold $ {T}_l $. For a negative opinion, the reputation value is $ {R_{j \to f}} $ less than the reputation threshold $ {T}_l $. The number of positive and negative recommendations is $ a $ and $ b $ respectively. The weighted aggregation of positive opinions is:
\begin{equation}
\label{equ2}
{P_{i \to f}} = \frac{{\sum\limits_{j = 1}^a {{C_{i \to j}} \cdot } {R_{i \to j}} \cdot {R_{j \to f}}}}{a}
\end{equation}
and the weighted aggregation of negative opinions is:
\begin{equation}
\label{equ3}
{N_{i \to f}} = \frac{{\sum\limits_{j = 1}^b {{C_{i \to j}} \cdot } {R_{i \to j}} \cdot {R_{j \to f}}}}{b}
\end{equation}
Based on the number of positive opinions and negative opinions, we can separately find their weights to find an indirect reputation value. The weights of positive and negative opinions are defined by Eq. \hyperref[equ4]{(4)} and Eq. \hyperref[equ5]{(5)}, respectively.

\begin{equation}
\label{equ4}
c = \frac{a}{{a + b}}
\end{equation}
\begin{equation}
\label{equ5}
d = \frac{b}{{a + b}}
\end{equation}
The \textit{indirect reputation value} of vehicle $ i $ to vehicle $ f $ is:
\begin{equation}
\label{equ6}
{{Ri}}{{{n}}_{i \to f}} = {c} \cdot {P_{i \to f}} - d \cdot {N_{i \to f}}
\end{equation}
\subsection{Feedback similarity}
 \label{sec5.2}
\par The TWSL model assumes that all vehicles provide honest ratings of other vehicles, but this assumption makes it difficult to identify malicious vehicles. Since a malicious vehicle may not only send fake messages but may also create malicious, inaccurate ratings. “Pretending-type” (P-type) malicious vehicles may send real messages before beginning to emit fake messages. Two honest (or malicious) vehicles are likely to be given similar ratings by the same group of vehicles that have interacted with them, but this is unlikely to be the case for an honest vehicle versus a malicious vehicle.

Here, we use the same group of vehicles $ Com(i, j) $ (e.g., vehicle $ q $) to indicate that have interacted with both vehicles $ i $ and $ j $, and calculate the feedback similarity of vehicle $ i $ and vehicle $ j $, thereby distinguishing P-type malicious vehicles from others. If the similarity of the feedback ratings of vehicle $ i $ and vehicle $ j $ is high, then vehicle $ j $ has a high reputation value to vehicle $ i $. The feedback formula of the overall rating of vehicle $ i $ to $ q $ is as follows (and is also applicable to the feedback of vehicle $ j $ to $ q $):

\begin{equation}
\label{equ7}
F(i,q) = \frac{{{f_1} \cdot \alpha - {f_2} \cdot \beta }}{{\alpha + \beta }}
\end{equation}
where $ \alpha,\beta $ are the number of positive ratings and negative ratings and their weights are $ {f_1},{f_2} $ respectively, as shown in Eq. \hyperref[equ8]{(8)} and Eq. \hyperref[equ9]{(9)}.
\begin{equation}
\label{equ8}
{f_1} = \frac{\alpha }{{\alpha  + \beta }}
\end{equation}
\begin{equation}
\label{equ9}
{f_2} = \frac{\beta }{{\alpha  + \beta }}
\end{equation}
\par We used the Weighted Euclidean Distance (WED) method to calculate the feedback similarity of vehicles. We determined a dispersion value based on the historical rating feedback of vehicle $ i $ and vehicle $ j $; smaller dispersion indicates greater similarity. The similarity of vehicle $  i  $ and vehicle $ j $ based on feedback is defined as \citep{liu2016machine}:
\begin{equation}
\label{equ10}
\begin{aligned}
&simf(i,j) = \\
&1 - \sqrt {\sum\nolimits_{q \in Com(i,j)} {Q(i,j,q) \cdot {{(F(i,q) - F(j,q))}^2}} } 
\end{aligned}
\end{equation}
where $ Q(i,j,q) $ represents the normalized weight of the influence of vehicle $ q $ on the similarity, and its standard deviation is $ Q'(i,j,q) $. The equations above were provided by \citep{liu2016machine}.

\par WED can be used to assign different weights to amplify the different ratings of vehicles $  i  $ and $  j  $ for the same group of participants. We utilized the exponential function of feedback similarity to define the \textit{local reputation confidence} $ {r_{i \to j}} $ (Eq. \hyperref[equ11]{(11)}), thereby preventing P-type vehicles accumulate deliberately high reputation values from certain honest vehicles. If there is no common interactive vehicle, the local reputation value weight is $ \theta $ $ \in  $ [0, 1].
\begin{equation}
\label{equ11}
{r_{i \to j}} = \mathop e\nolimits^{(1 - 1/simf(i,j))} 
\end{equation}
When the feedback similarity of the vehicle is high, $ {r_{i \to j}} $ is also high (and vice versa).
\subsection{Final reputation value}
 \label{sec5.3}
 The \textit{final reputation value} $ {{Rfi}}{{{n}}_{i \to f}} $ can be divided into the following situations:
\begin{enumerate}[1)]
	\item Vehicle $ i $ and vehicle $ f $ have never interacted and have no recommendations. The initial local reputation value from  $  i  $ to $ f $ is $ \gamma $ $ \in $ [0, 1] and the final reputation value is:
	\begin{equation}
	\label{equ12}
	{{Rfi}}{{{n}}_{i \to f}} = {{{r}}_{i \to f}} \cdot \gamma
	\end{equation}
	\item Vehicle $ i $ and vehicle $ f $ have not established interactions before, but neighbor vehicles $ j $ has a recommendation to $ f $. The local reputation value is $ \eta $ $ \in $ [0, 1] and the final reputation value of $ i $ to $  f  $ is:
	\begin{equation}
	\label{equ13}
	{{Rfi}}{{{n}}_{i \to f}} = {{{r}}_{i \to f}} \cdot \eta + {(1 - }{{{r}}_{i \to f}}) \cdot Ri{n_{i \to f}}
	\end{equation}
where $ Ri{n_{i \to f}} $ is given by Eq. \hyperref[equ2]{(2)}, Eq. \hyperref[equ3]{(3)} and \\ Eq. \hyperref[equ6]{(6)}.
	\item Vehicle $ i $ has established interaction with vehicle $ f $, but no neighboring vehicle has a recommendation to $ f $. The final reputation value of $ i $ to $ f $ is:
	\begin{equation}
	\label{equ14}
	{{Rfi}}{{{n}}_{i \to f}} = {{{r}}_{i \to f}} \cdot {R_{i \to f}}
	\end{equation}
	\item Vehicle $ i $ has established interaction with vehicle $ f $, and the neighboring vehicle has a recommendation to $ f $ and the final reputation value of $ i $ to $ f $ is:
	\begin{equation}
	\label{equ15}
	Rfi{n_{i \to f}} = {r_{i \to f}} \cdot {R_{i \to f}} + (1 - {r_{i \to f}}) \cdot Ri{n_{i \to f}}
	\end{equation}
\end{enumerate} 
\begin{figure}[ht]
	\centering
	\subfigure[Reputation management.]{
		\includegraphics[scale=0.23]{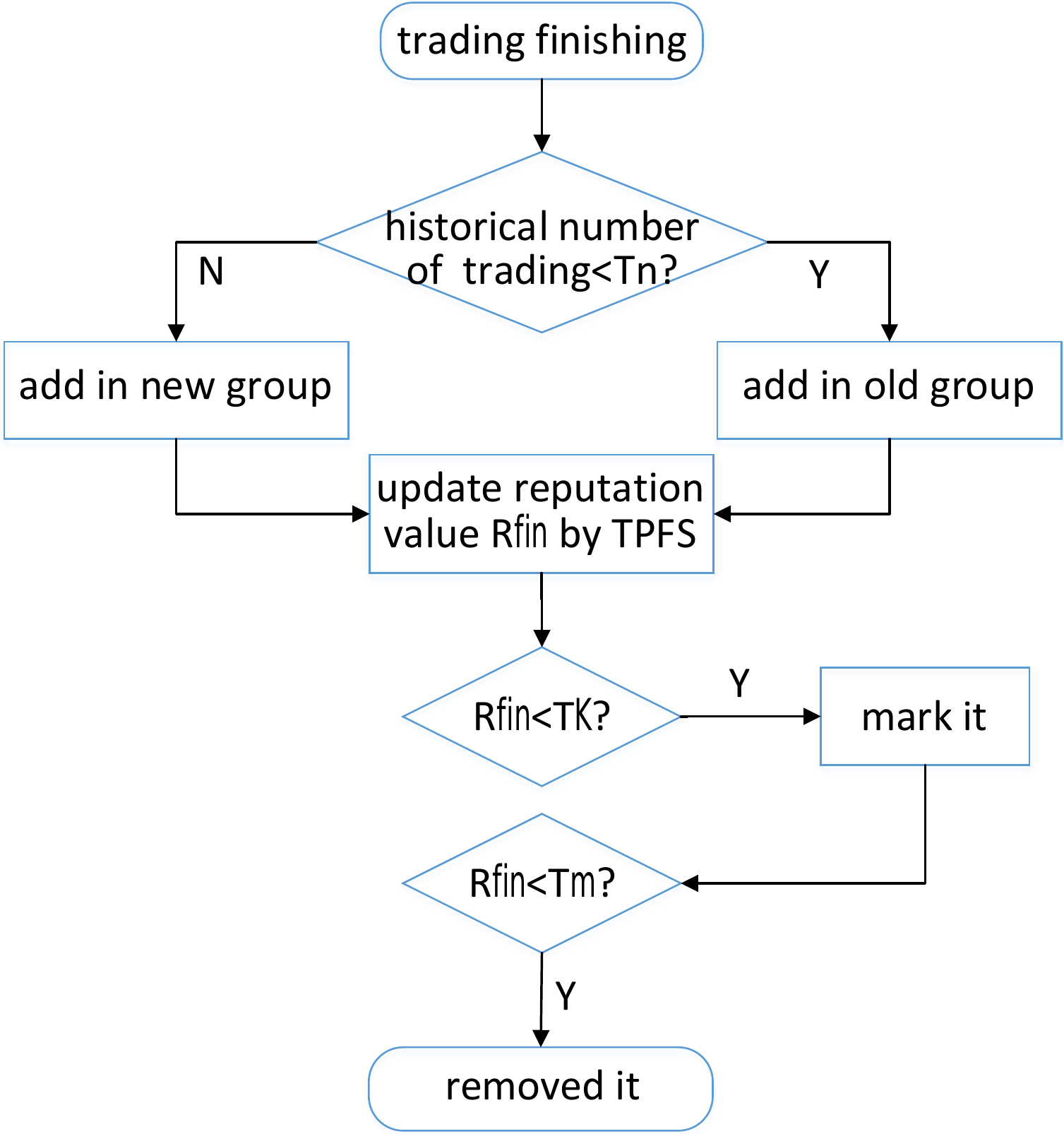}

	}
	\quad
	\subfigure[Fair server selection.]{
		\includegraphics[scale=0.22]{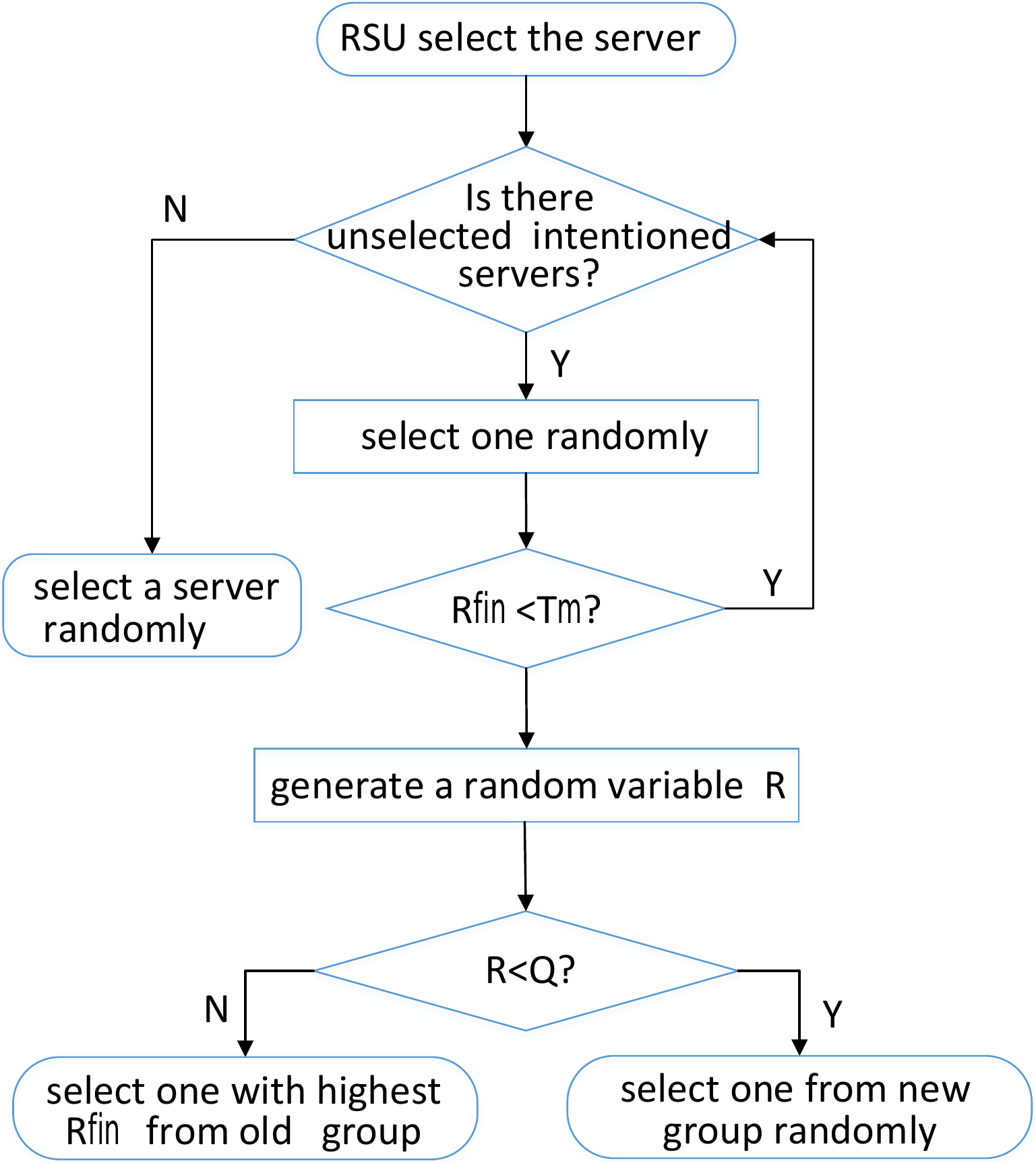}

	}
	\caption{Fairness mechanism of TPFS.}
	\label{fig4}
\end{figure}
It is worth mentioning that in the process of updating the reputation value based on the chaincode, and the reputation value can be traced through the transaction records of the block. In addition, if the reputation value of a server is lower than the suggested service threshold $ T_k $, it is marked to remind the requester to proceed with caution. If a server’s reputation value is lower than the allowable threshold $ T_m $, its identity certificate is revoked – the consortium removes it and does not permit it again in the future. A flowchart of reputation management processes is given in \hyperref[fig4]{Fig. 4(a)}. We also divided the intentioned servers into two groups, one new and one old, based on the historical number of trading threshold $ T_n $ to enhance the fairness of the server selection process. This provides service opportunities for vehicles newly joining the consortium. 

RSU generates a random number $ R $ to compare with the value of $ Q \in $ [0, 1], which is a threshold specified by chaincode. \hyperref[fig4]{Fig. 4(b)} shows the detailed flowchart of the server selection. There may also be all intentioned servers whose reputation value is relatively low. In this case, servers are selected randomly. (Of course, this is rare.)

\subsection{TPFS model evaluation }
\label{sec5.4}
\par We conducted a series of experiments through Matlab to evaluate the performance of TPFS and to compare it against TWSL. Parameters in the experiment included $ \gamma $=0.2, $ \eta $=0.2, and $ \theta $=0.7 (Section~\ref{sec5.3}).

\par Firstly, we analyzed the changes in reputation values over time based on different models as vehicles interact. Most services in the IoV have timeliness requirements. We considered segmented interactions in different situations to determine whether the reputation value can be updated in time. \citet{kang2018blockchain} set the reputation value update time to 1 minute in their work; similarly, we assumed in this experiment that vehicle $  i  $ and vehicle $  j  $ interact once per minute for a total of 100 interactions with 100 minutes. Vehicle $  j  $ sends real messages in [0, 50] minutes and sends fake messages in [51, 80] minutes. Negative ratings have a greater impact on vehicle reputation than positive ratings, so the reputation declines rapidly in the latter half of this process. In [81, 100] minutes, vehicle $  i  $ and vehicle $  j  $ do not interact. 
\begin{figure}[ht]
	\centering
	\includegraphics[scale=0.6]{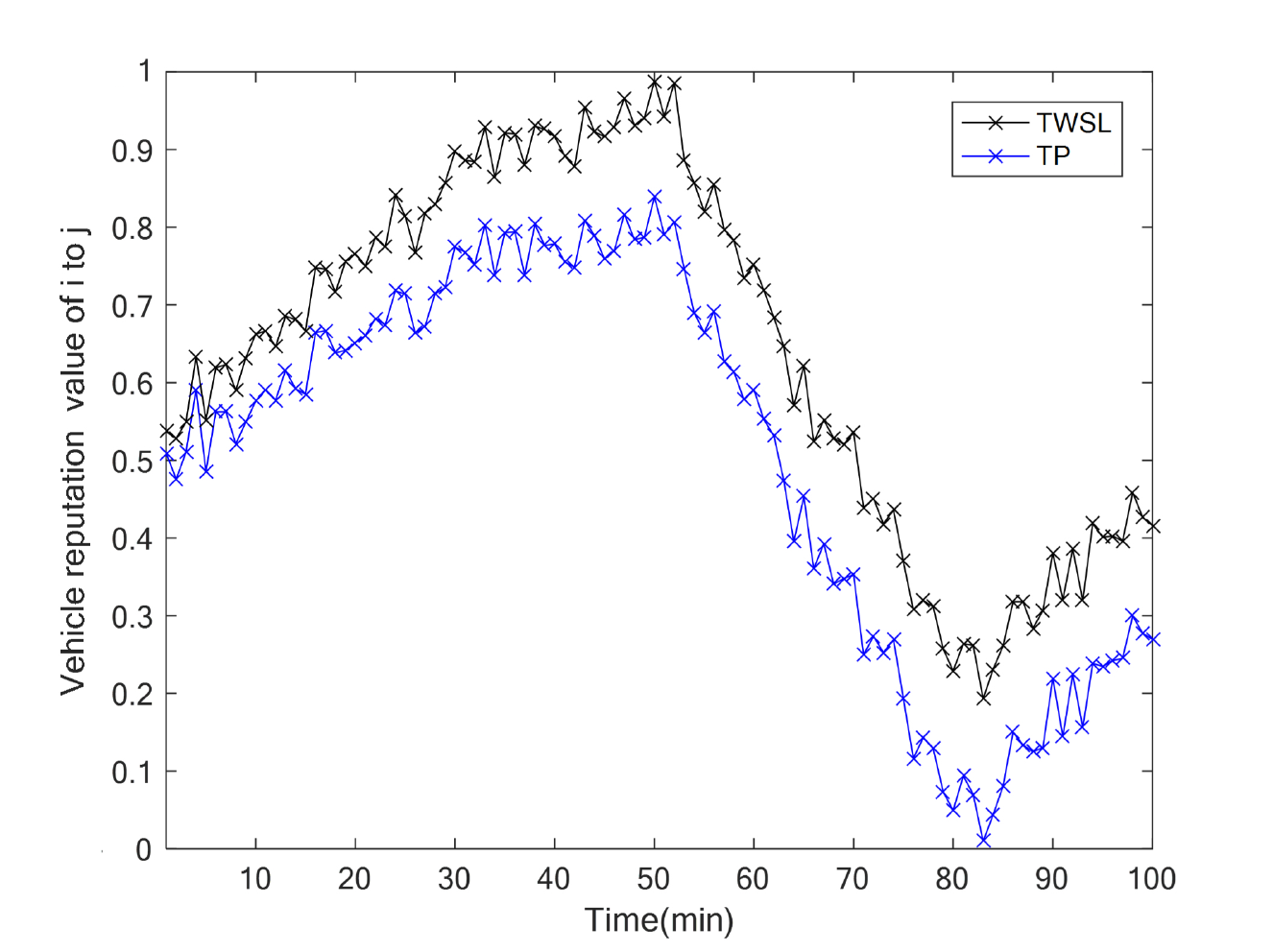}
	\caption{Reputation changes of vehicles over time.}
	\label{fig5}
\end{figure}

At this time, the impact of previous ratings on vehicle reputation value gradually decreases while the influence of neighbor vehicle recommendation gradually increases.
\hyperref[fig5]{Fig. 5} shows the change in the reputation value of the vehicle $  i  $ to $  j  $ over time. Because TP considers recommendations from malicious vehicles, the reputation value it obtains is lower than that obtained by TWSL.

\begin{figure}[ht]
	\centering
	\includegraphics[scale=0.6]{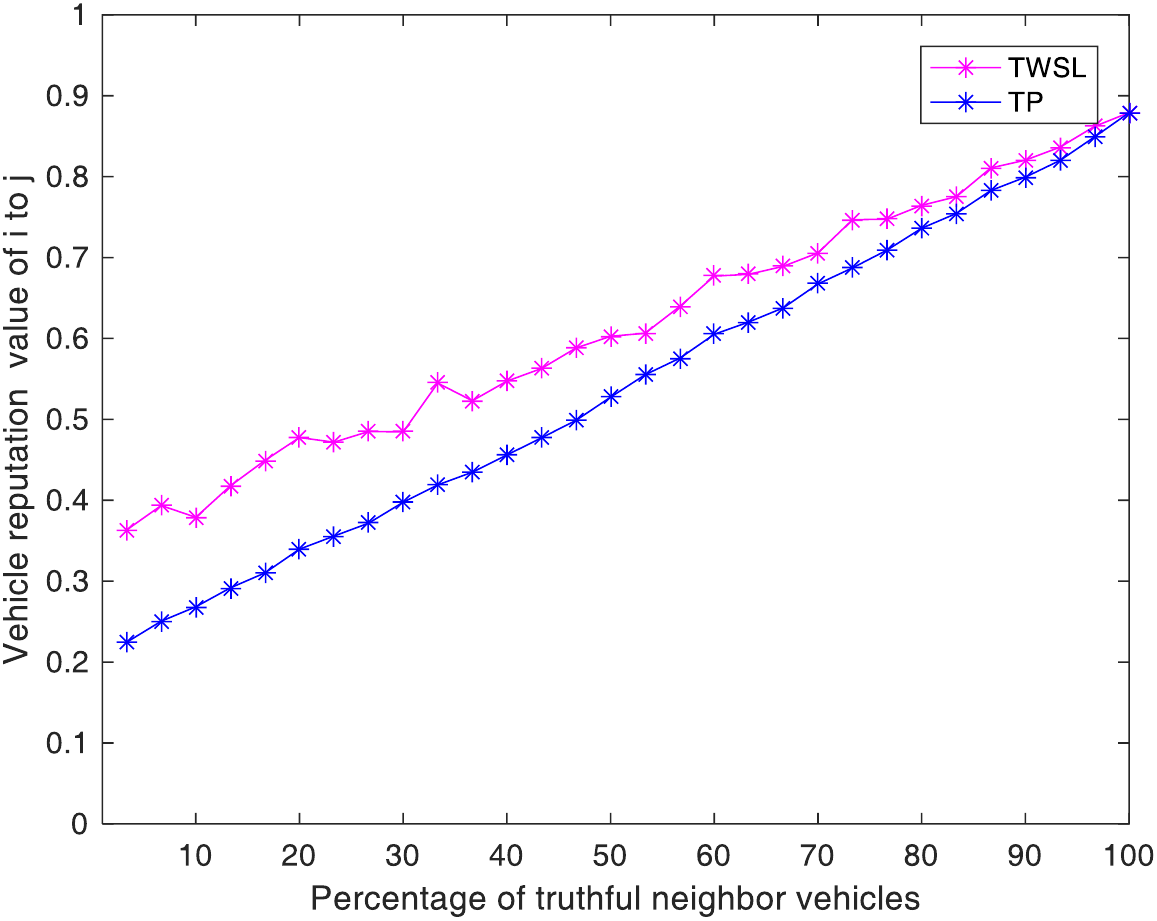}
	\caption{Relationship between percentage of truthful neighbor vehicles and the vehicle reputation value.}
	\label{fig6}
\end{figure}
\par We also analyzed the impact of trusted neighbor recommendations on reputation. We set a total of 30 neighboring vehicles. In the case of no interaction between vehicle $  i  $ and vehicle $ j $, the reputation value of vehicle $  i  $ to vehicle $ j $ originates entirely from “trust propagation”, that is, the recommendations of other neighbor vehicles. \hyperref[fig6]{Fig. 6} shows the relationship between the percentage of truthful neighbor vehicles to all neighbor vehicles and the reputation of vehicle $  i  $ to $ j $.
The reputation value of vehicle $  i  $ to vehicle $ j $, obtained by TP, is lower when there are fewer trusted vehicles providing recommendations. This is because TWSL ignores the fact that malicious vehicles may lead to malicious recommendations. TP first evaluates the credibility of the recommended vehicle and can reject any unreliable recommendation provided by a malicious vehicle. 
\begin{figure}[htbp]
	\centering
	\includegraphics[scale=0.6]{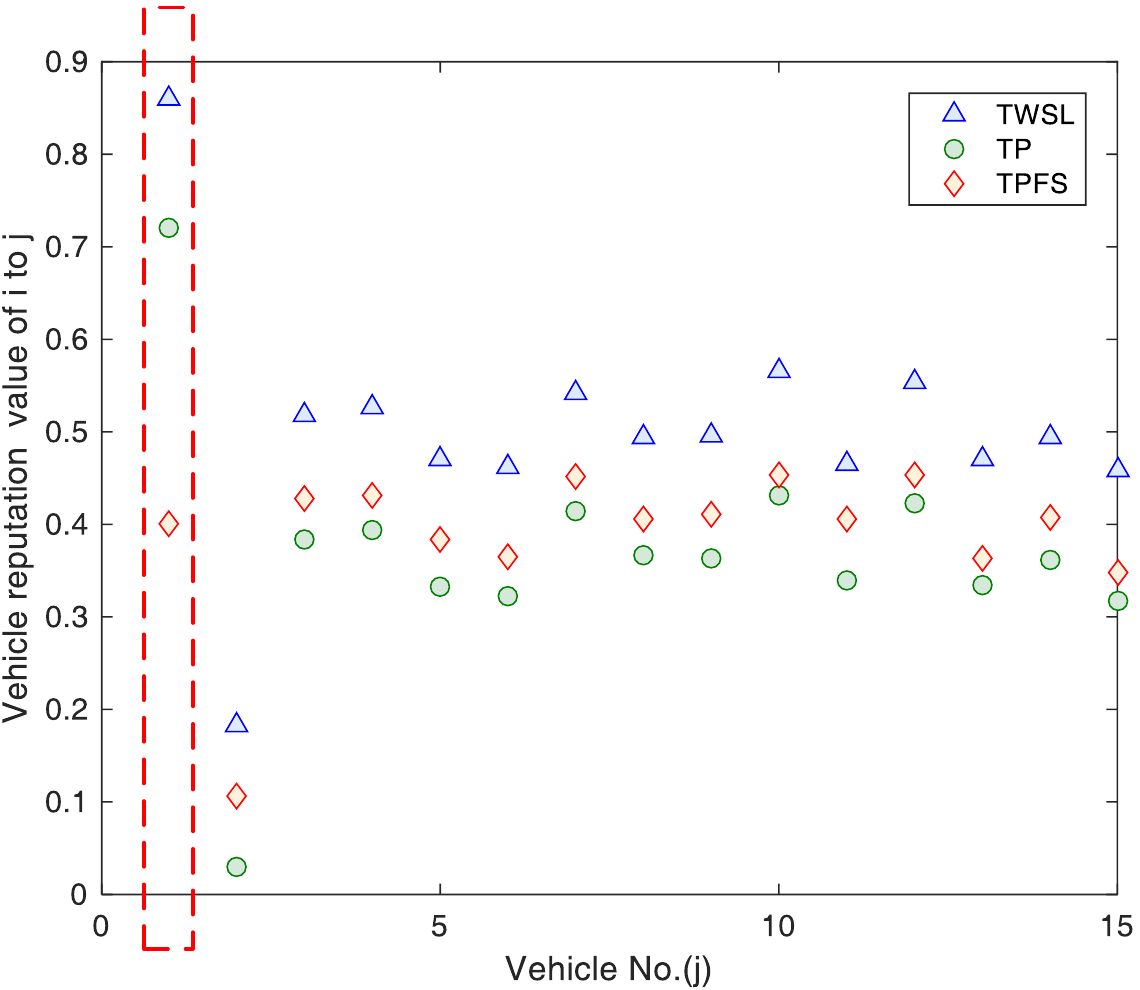}
	\caption{Reputation value distribution of all vehicles with\\ P-type.}
	\label{fig7}
\end{figure}

\par It is also important to consider that malicious vehicles may have unstable behavior. For the next experiment, we set 15 distributed vehicles as servers randomly. As shown in \hyperref[fig7]{Fig. 7}, No. 1 is a malicious vehicle that sends numerous real messages to confuse vehicle $ i $. It attempts to conduct malicious attacks after gaining a high reputation value. After 100 interactions, we obtained the reputation value of 15 vehicles as shown in \hyperref[fig7]{Fig. 7}. The reputation values calculated based on TWSL and TP are higher than those obtained by TPFS, because TPFS reduces its reputation value by identifying the malicious vehicle.

\section{Performance evaluation}
\label{sec6}
We next attempted to optimize the proposed RC-chain configuration, by applying queueing theory to quantitatively evaluate the impact of transaction arrival rate and \textit{batch size} (i. e., maximum number of transactions in a block) on transaction confirmation time. 
\subsection{Queue model of RC-chain}
\label{sec6.1}
We designed a queueing system with three different service stages, that express the three-stage consensus process and the building of a new block. By analyzing the queueing model, we obtained three key performance measures: 1) the \textit{average number of transactions} in system; 2) the \textit{average transaction confirmation time}, and 3) the \textit{average transaction throughput}. 

\begin{figure}[ht]
	\centering
	\includegraphics[scale=0.6]{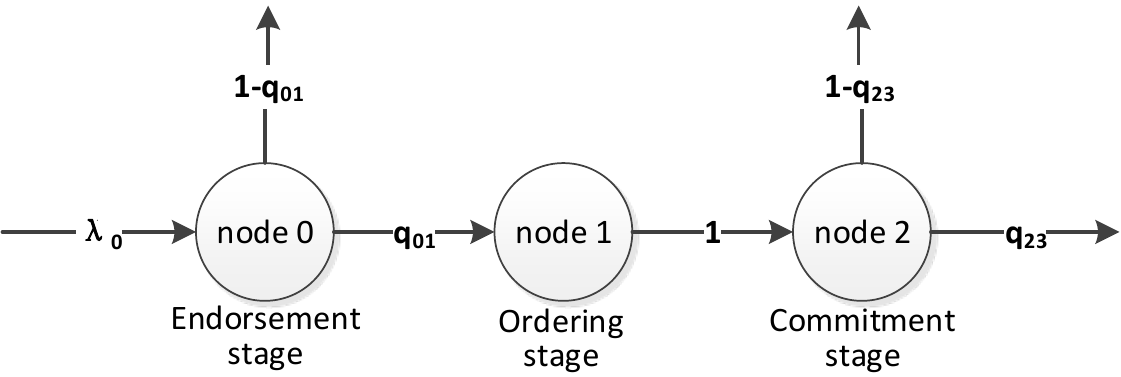}
	\caption{Queueing network model.}
	\label{fig8}
\end{figure}

The notations used in this section, which are related to the RC-chain consensus mechanism (Section ~\ref{sec4}), are summarized in \hyperref[tab2]{Table 2}. A diagram of the model is given in \hyperref[fig8]{Fig. 8}.

We constructed a three-node open queueing network \citep{Hayes2004Modeling}. The endorsement stage is node 0, the ordering stage is node 1, and the commitment stage is node 2. Assuming that the transaction arrival from clients is a Poisson process, the arrival rate can be denoted as $\lambda_{0}$ and the external arrival to each node is represented by the vector $\boldsymbol{\lambda}$ = [$\lambda_{0}$, 0, 0].

\begin{table}
	\label{tab2}
	\begin{tabular}{ll}
		\multicolumn{2}{l}{\small{\textbf{Table 2}}}\\
		\multicolumn{2}{l}{\small{Notaions}}\\
		\toprule
		Symbols & Meanings \\
		\midrule
		$ q_{ji} $ & {\tabincell{l}{-probability of a transaction being \\routed from node $ j $ to node $ i $}} \\
		$\lambda_i$ & \multicolumn{1}{l}{-arrival rate of transactions to node $ i $}\\
		$ \Lambda_{i} $ & \multicolumn{1}{l}{-total transactions arrival rate to node $ i $}\\
		$\mu_i $ &\multicolumn{1}{l}{-service rate of transactions to node $ i $}\\
		$ R_i $ & \multicolumn{1}{l}{-service intensity of node $  i  $}\\
		$ P(k_i) $ & {\tabincell{l}{-probability that there are $ k_i $ \\transactions in node $ i $}}\\
		$ P(k_0,k_1,k_2) $ & {\tabincell{l}{-state probability that there \\are $ k_0 $, $ k_1 $, $ k_2 $ transactions in \\node 0, 1, 2, respectively}}\\
		$ \bar N_i $ & {\tabincell{l}{-average number of transactions\\ in node $ i $}}\\
		$ \bar D_{{i}} $ & {\tabincell{l}{-average delay of a transaction \\through node $  i  $}}\\
		$ \bar D $ & {\tabincell{l}{-average transaction confirmation time}}\\
		$ H $ & {\tabincell{l}{-average transaction throughout}}\\
		$ M $ & {\tabincell{l}{-\textit{batch size} (i. e., maximum number of\\ transactions in a block)}}\\
		\bottomrule
	\end{tabular}
\end{table}

In RC-chain, the transactions that do not meet the endorsement policy or fail the verification are not be updated to the ledger. Such invalid transactions are considered to be leaving the queueing network. The probability that a transaction is routed from node $ j $ to node $ i $ is given by $ q_{ji} $. The probabilistic routing matrix is:

\begin{equation}
	\label{equ16}
	Q =
	\begin{bmatrix}
		0 & q_{01} & 0 \\
		0 & 0 & 1 \\
		0 & 0 & 0
	\end{bmatrix}
\end{equation}
The total traffic into each node of the the network is:
\begin{equation}
	\label{equ17}
	{\Lambda_{i}} = {\lambda_i} + \sum\limits_{j = 0}^2 {{q_{ji}}} \cdot {\Lambda _j}\quad i = 0,1,2 
\end{equation}
As per the probabilistic routing matrix $ Q $ and the external arrival $\boldsymbol{\lambda}$,
\begin{equation}
	\label{equ18}
	\begin{split}
		{\Lambda _0} &= {\lambda _0}\\
		{\Lambda _1} &= {\Lambda _2} = {q_{01}} \cdot {\lambda _0}   
	\end{split}  
\end{equation}
Assuming that each node in queueing network has unlimited storage space, and the transaction service time in each node is exponentially distributed with average rate $\mu_i $, $ i $ = 0, 1, 2.

At any consensus stage in the RC-chain, multiple servers serve the same transaction independently and simultaneously. For example, in the endorsement stage, multiple endorsing peers simultaneously validate the same transaction proposals. Therefore, we assume that there is one server at each node in the queueing network.

In Hyperledger Fabric, there are two conditions for orderers to perform block packaging: the number of transactions waiting packaged reaches the \textit{batch size}, or these transaction waiting time reaches maximum elapsed duration (i.e., batch timeout). When either condition is met, the orderers immediately package the transactions into blocks. Here, we take the first condition as the model setting and let $ M $ denote the \textit{batch size}. Because the transaction arrival rate to the ordering stage is $\Lambda_1 = q_{01}\cdot\lambda _0$, the average arrival time interval for a transaction is $ 1/\Lambda_1 $. For $ M $ transactions, the time to wait for packaging after the transaction arrives is (0, $M/\Lambda_1 $). The service rate of the orderers is given by

\begin{equation}
	\label{equ19}
	{\mu_1} = 2{\Lambda _1}/M 
\end{equation}
where the service rate of the committing peer is assumed to be equal to that of the endorsing peers.

Let $ Q1 $, $ Q2 $, and $ Q3 $  denote the queue length in node 0, 1, and 2 in the steady state, respectively. At this point, the model is identical to the 3-node open Jackson network. The state of the system is defined by the number of transactions in each of the nodes: $ k_0 $, $ k_1 $, $ k_2 $. By strictly applying the derivation given by \citep{Hayes2004Modeling}, we have

\begin{equation}
	\label{equ20}
	\begin{aligned}
		&P(Q_0 = {k_0},Q_1 = {k_1},Q_2 = {k_2}) = \\
		&P({k_0},{k_1},{k_2}) = \prod\limits_{i = 0}^2 {(1 - {R_i}){R_i}^{{k_i}}}  \\
	\end{aligned}
\end{equation}
where
\begin{equation}
	\label{equ21}
	{R_{{i}}} = {\Lambda _i}/{\mu _i} 
\end{equation}
represents the service intensity of each node. The steady state condition of the system is $ 	{R_{{i}}}/{S_i} = {R_i} < 1 $
\par According to Eq. \hyperref[equ18]{(18)} and Eq. \hyperref[equ21]{(21)},
\begin{equation}
	\label{equ23}
	\begin{split}
		{R_0} &= {{{\lambda _0}} \mathord{\left/
				{\vphantom {{{\lambda _0}} {{\mu _0}}}} \right.
				\kern-\nulldelimiterspace} {{\mu _0}}}\\
		{R_1} &= {{{q_{01}} \cdot {\lambda _0}} \mathord{\left/
				{\vphantom {{{q_{01}} \cdot {\lambda _0}} {{\mu _1}}}} \right.
				\kern-\nulldelimiterspace} {{\mu _1}}}\\
		{R_2} &= {{{q_{01}} \cdot {\lambda _0}} \mathord{\left/
				{\vphantom {{{q_{01}} \cdot {\lambda _0}} {{\mu _2}}}} \right.
				\kern-\nulldelimiterspace} {{\mu _2}}}
	\end{split}
\end{equation}
According to Law of Total Probability, the state equation of each node is:
\begin{equation}
	\label{equ24}
	P(Q_0) = P({k_0}) = \sum\limits_{{k_2} = 0}^\infty  {\sum\limits_{{k_1} = 0}^\infty  {P({k_0},{k_1},{k_2})} } = (1 - {R_0}){R_0}^{{k_0}}
\end{equation}
\begin{equation}
	\label{equ25}
	P(Q_1) = P({k_1}) = \sum\limits_{{k_2} = 0}^\infty  {\sum\limits_{{k_0} = 0}^\infty  {P({k_0},{k_1},{k_2})} } = (1 - {R_1}){R_1}^{{k_1}}
\end{equation}
\begin{equation}
	\label{equ26}
	P(Q_2) = P({k_2}) = \sum\limits_{{k_1} = 0}^\infty \\
	{\sum\limits_{{k_0} = 0}^\infty  {P({k_0},{k_1},{k_2}) = (1 - {R_2}){R_2}^{{k_2}}} } 
\end{equation}
This system satisfies Jackson’s theorem. The average number of transactions in each node is:
\begin{equation}
	\label{equ27}
	{\bar N_i} = \sum\limits_{{k_i} = 0}^\infty  {{k_i}P({k_i})}  = \frac{{{R_i}}}{{(1 - {R_i})}} \quad {i} = 0,1,2
\end{equation}
The \textit{average number of transactions} in system is:
\begin{equation}
\label{equ28}
	\begin{aligned}
		&\bar N = \sum\limits_{{k_0},{k_1},{k_2} = 0}^\infty  {({k_0} + {k_1} + {k_2})}P({k_0},{k_1},{k_2}) = \\
		&\frac{{{R_0}}}{{(1 - {R_0})}} + \frac{{{R_1}}}{{(1 - {R_1})}} + \frac{{{R_2}}}{{(1 - {R_2})}} \\
	\end{aligned}
\end{equation}
As per Little’s theorem, the average delay of a transaction through node $ i $ is:
\begin{equation}
\label{equ29}
	{\bar D_{{i}}} = \frac{{{{{{\bar N}}}_{{i}}}}}{{{{\Lambda} _{{i}}}}} = \frac{{{R_{{i}}}}}{{(1 - {R_i}) \cdot {\Lambda _i}}}
\end{equation}
Let $ \bar D $ denote the \textit{average transaction confirmation time}, the time interval from the arrival time point of a transaction to its departure. Be noted that the transaction confirmation time here is calculated from the arrival endorsement stage of the transaction, and the time from the client to the endorsement peers is ignored for simplicity of analysis. Then,
\begin{equation}
\label{equ30}
	\bar D = \sum\limits_{{{i}} = 0}^2 {{{\bar D}_i}}
\end{equation}

The \textit{average transaction throughput $ H $} is defined as the rate at which verified transactions are committed by the blockchain across the entire network, that is, the number of transactions that are updated in the ledger per unit time. $ H $ is given by

\begin{equation}
\label{equ31}
	H = \frac{{{{\bar N}_2} \cdot {q_{23}}}}{{\bar D}} = \frac{{{R_2} \cdot {q_{23}}}}{{(1 - {R_2}) \cdot \bar D}}
\end{equation}
\subsection{Numerical analysis}
\label{sec6.2}
Transaction confirmation time is a key factor for evaluation blockchain-based VCS platform, and the trust information dissemination in vehicular network is time-sensitive. Therefore, we discuss the time performance in transaction confirmation.

Through the numerical calculation on Matlab, we analyze with average transaction confirmation time and average transaction delay at orderer of RC-chain. The results show that the performance of RC-chain based Hyperledger Fabric occupies a clear advantage. 

We take some common parameters: transaction arrival rate $\lambda_0\in$ [10, 110], \textit{batch size} $ M $=10, 50, 100, $ q_{01} $=0.9, $ q_{23} $=0.95. The service rate of both the endorsing peers and the committing peers are 150. 
\begin{figure}[ht]
	\centering
	\includegraphics[scale=0.36]{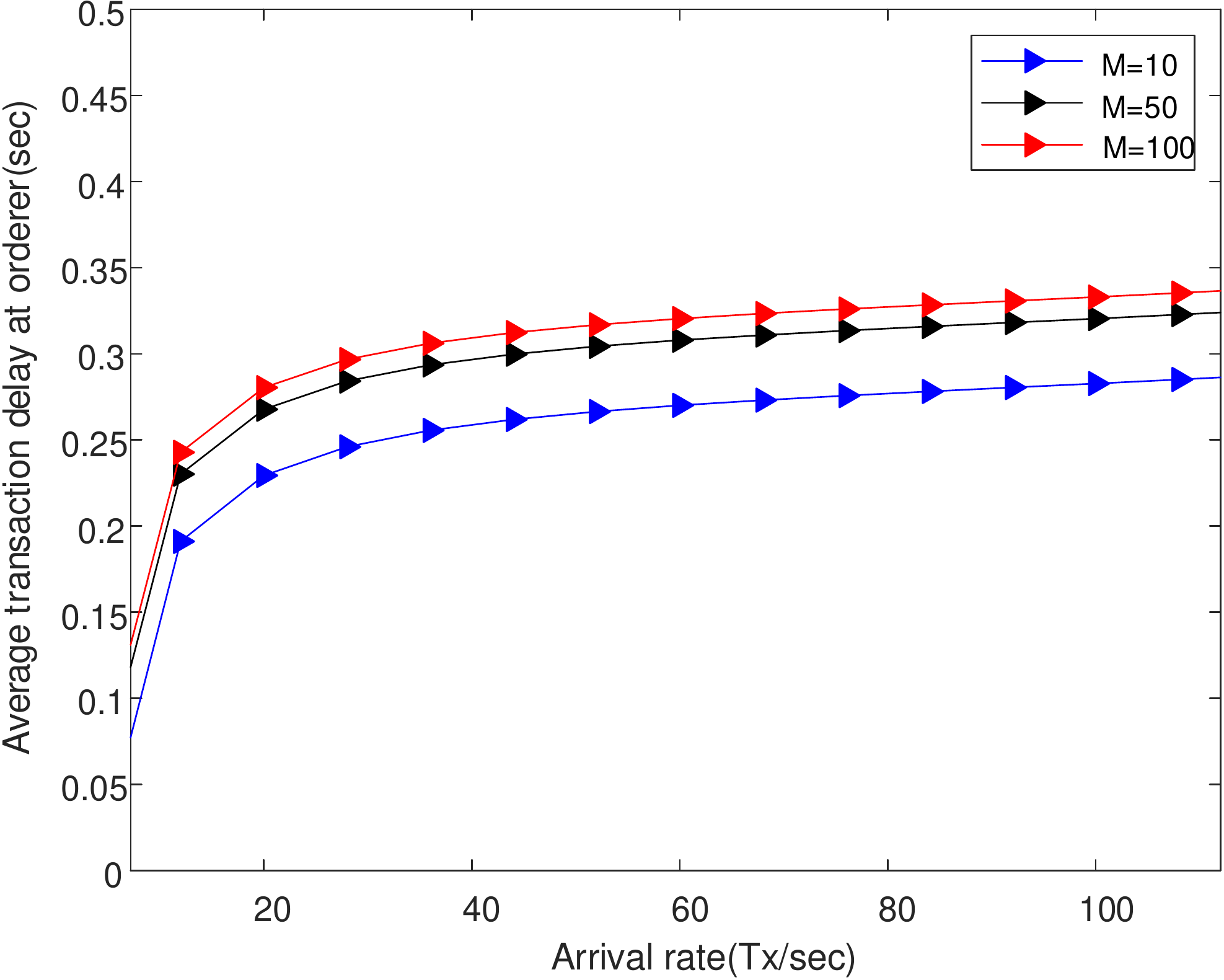}
	\caption{The average delay of transaction at orderers.}
	\label{fig9}
\end{figure}
\begin{figure}[ht]
	\centering
	\includegraphics[scale=0.28]{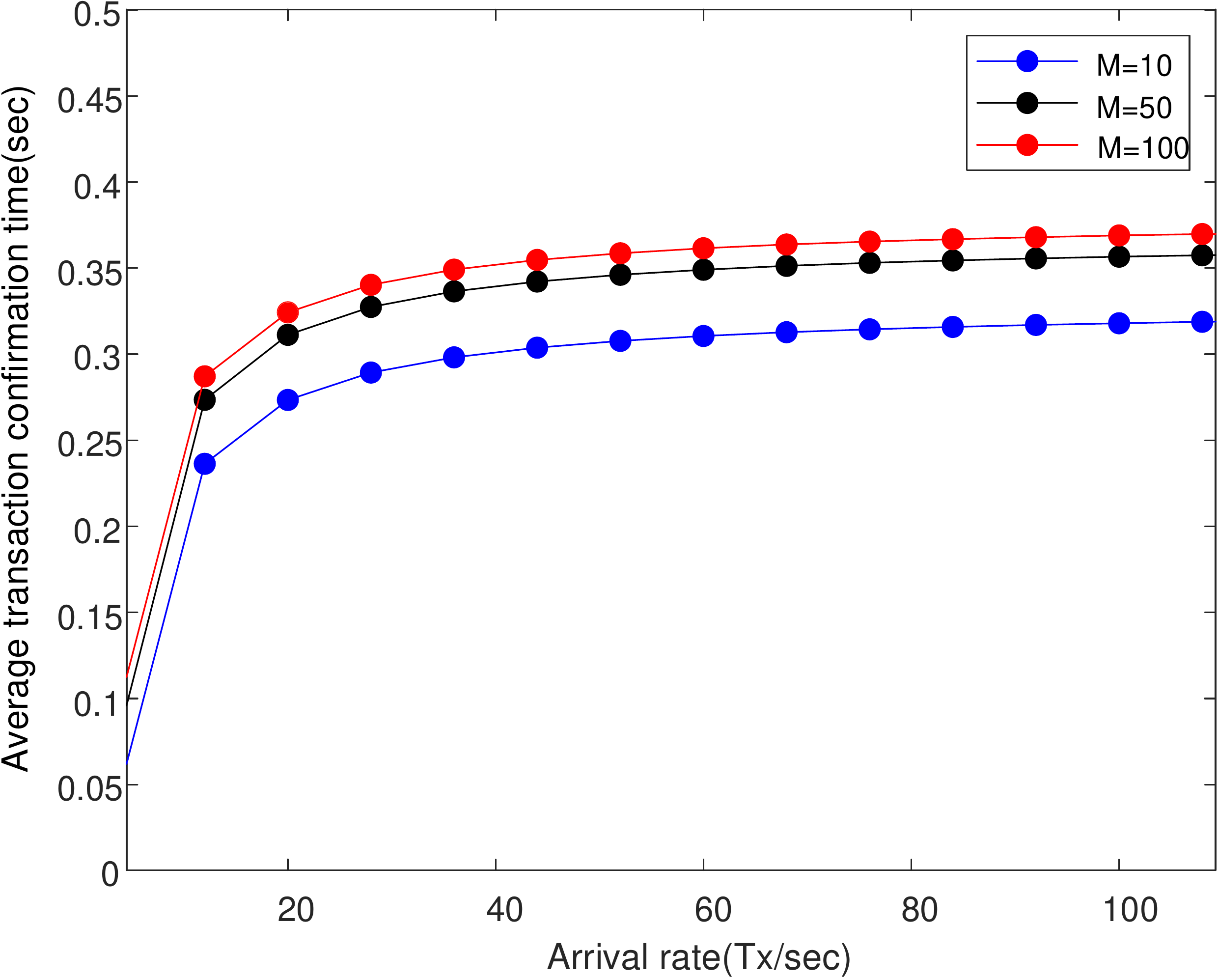}
	\caption{The average transaction confirmation time.}
	\label{fig10}
\end{figure}
\par It should be noted that the ordering stage has a great impact on the efficiency of block generation. \hyperref[fig9]{Fig. 9} shows the relationship between average transaction delay at orderer and the transaction arrival rate. It can be seen, as the transaction arrival rate increases, the average delay of the transaction at orderer gradually increases. When the arrival rate increases to 40, the value converge smoothly. The change of average transaction confirmation time with arrival rate has the same trend, as shown in \hyperref[fig10]{Fig. 10}. In addition, the smaller the \textit{batch size}, that is, the higher the service rate of orderers, the shorter the average transaction delay at orderer, and so does the average transaction confirmation time. Take \hyperref[fig10]{Fig. 10} as an example, when $ M $ is 100, the average delay is about 500 microseconds higher than $ M $ is 10, and about 108 microseconds higher than $ M $ is 50, which provides an effective basis for the next experiment deployment.

\section{Experiment results}
 \label{sec7}
We next analyzed the safety and trust goals that are achievable by RC-chain. We constructed a VCS network, wrote smart contract, and implemented RC-chain using the configuration determined by the above model parameters. The running results were compared with the theoretical results of the queueing model and an existing similar scheme. 
\begin{table*}[bp]
	\label{tab3}
	\centering
	\begin{tabular*}{16cm}{l|cccccccccc}  
		\multicolumn{3}{l}{\small{\textbf{Table 3}}}\\
		\multicolumn{3}{l}{\small{Transaction confirmation time comparison}}\\
		\hline  
		block No. & 1 &2&3&4 & 5 &6&7&8& 9 &10\\  
		\hline  
		arrival rate(Tx/sec) & 37.29 &34.09&32.91&37.95 & 37.20 &36.12&37.66&37.33 & 39.39 &36.95\\ 
		\hline   
		theoretical value(sec) & 0.299 &0.296&0.256&0.299&0.298&0.297&0.299&0.299& 0.300&0.298\\ 
		\hline  
		actual value(sec)&0.303&0.306&0.312&0.302&0.315&0.309&0.305&0.304&0.303 &0.311\\  
		\hline 
		deviation(sec)& 0.004 &0.010&0.056&0.003 &0.017 &0.012&0.006&0.005 & 0.003 &0.013\\ 
		\hline  
	\end{tabular*}  
\end{table*} 
\begin{figure*}[ht]
	\centering
	\includegraphics[scale=0.6]{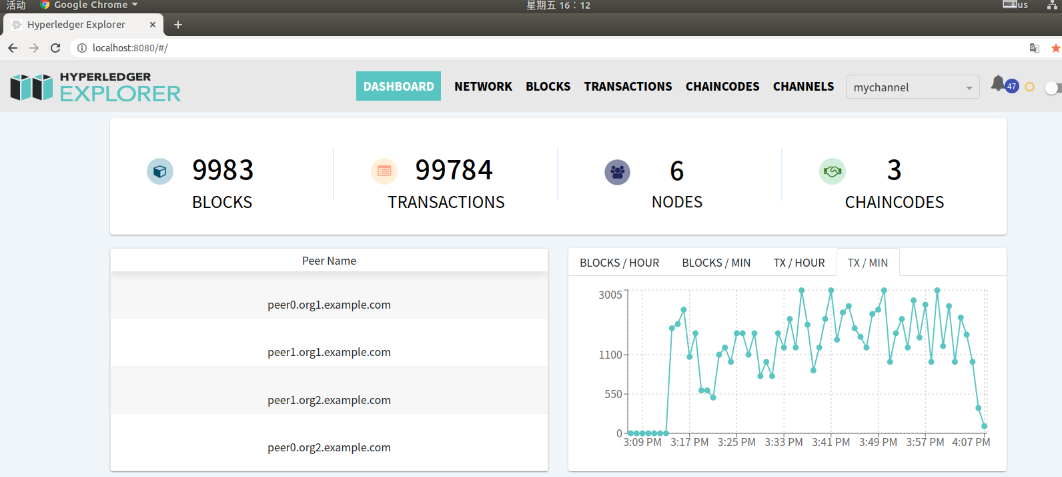}
	\caption{RC-chain implementation.}
	\label{fig11}
\end{figure*}
\subsection{Safe and trust gaols}
\label{sec7.1}
\par The RC-chain design combined with the consortium blockchain Hyperledger Fabric platform can achieve the following security goals.
\par \textbf{Distributed solution}: RC-chain is a distributed VCS solution. It does not rely on a unified manager during the transaction process. It is difficult for an attacker to target all nodes in the network, which enhances its fairness and security while reducing the possibility of the entire network being threatened by a single point attack.
\par \textbf{Permissioned identity}: The CA utilizes an elliptic curve digital signature algorithm to generate keys and certificates for the entity. The identity information is bound to its registration information and the physical address of the device, which uniquely identify this entity. Entities with malicious historical behavior have a low reputation value and cannot forge identities to join the network.
\par \textbf{Privacy protection}: The channel mechanism provides transaction privacy and confidentiality for a specific subset of network members. All data on the channel, including transactions, members, and channel information, is invisible to members outside the channel, which preserves privacy.
\par \textbf{Reliable reputation value management--TPFS model}: The legal identity represents the historical behavior of the entity. Its reputation value is updated timely based on current behavior. By degrading the reputation value of those service participants with untruthful behavior, adversarial or irresponsible behavior can be prevented.
\par \textbf{Transparent transaction process}: Participants, be they requesters, servers, or the RSU, upload the main steps of the transaction to the ledger. Nodes with copies of the ledger in the same channel can review and supervise others. If the node is malicious, it is easy to be found and held accountable. An entity with malicious behavior risks being forced out of the network due to a dropped reputation value. Thus, RC-chain can effectively prevent malicious behavior.
\par \textbf{Three-stage consensus to achieve BFT}: The transaction can only move into the subsequent consensus stage after sufficient endorsement signatures have been gathered in the endorsement stage. Illegal transactions cannot secure this sufficient quantity of endorsement signatures, thereby reducing attacks on the network. CFT is achieved via the second-stage Kafka-based ordering service. The final step is the verification process, which eliminates any double spending phenomenon and reviews transactions again. RC-chain implements BFT via this three-stage consensus.

\subsection{RC-chain implementation}
By using the Hyperledger Fabric running on Linux, we conducted another RC-chain experiment based on the results obtained from the above queueing theory model. We deployed RC-chain on a machine with Intel(R) Core(TM) i5-6300HQ CPU @ 2.30GHz and 6GB RAM running Ubuntu 18.04.4 LTS. The VCS network we built for this purpose contains three organizations, each one consisting of two peers and one orderer. The chaincode was written in GO language with the main functions of user registration, information query, data index acquisition, broadcast message, reputation value update, and server selection.
  \label{sec7.2}
According to the conclusion presented in Section~\ref{sec6.2}, we set \textit{batch size} $ M $ to 10, with 3 clients submitting transactions randomly in parallel at an average arrival rate of 40. As shown in \hyperref[fig11]{Fig. 11}, we randomly selected trading within 52 minutes and observed it through Hyperledger Explorer. A total of 99784 transactions and 9983 blocks were generated during this period with an average throughput of 32 Tx/s. The throughput reached a maximum of 50 Tx/s and the system was operating in a stable state.

We randomly selected 10 consecutive blocks, calculated and analyzed the actual transaction confirmation time in Hyperledger Fabric, and compared it with the theoretical transaction confirmation time obtained by the queueing model. The results are shown in \hyperref[tab3]{Table 3}. Because the queueing model does not consider propagation delay or other certain factors, the theoretical value is smaller than the actual value on average; however, the average deviation is between them is small enough to consider our scheme realistic and feasible.

We also compared the transaction confirmation time with the framework QcFND in IoV-based Fabric \citep{xiao2020edge}. We selected 30 blocks. Similar to the QcFND scheme, we selected five committing peers and calculated the average transaction confirmation time on five nodes with a \textit{batch size} of 10 and batch timeout of 2s, with negligible time from clients to the endorsing peers. The average transaction confirmation time of 30 blocks was 0.312s in this case and the average transaction confirmation time of 52 minutes period was 0.315s, so we were able to consider the partial representing approximately the whole. As shown in \hyperref[fig1]{Fig. 12}, the average transaction time of the QcFND scheme was 2.194s and the average transaction time of RC-chain was 0.315s in this test; the transaction confirmation time of each of the 30 blocks is also shown in the figure.
\begin{figure}[htpb]
	\centering
	\includegraphics[scale=0.6]{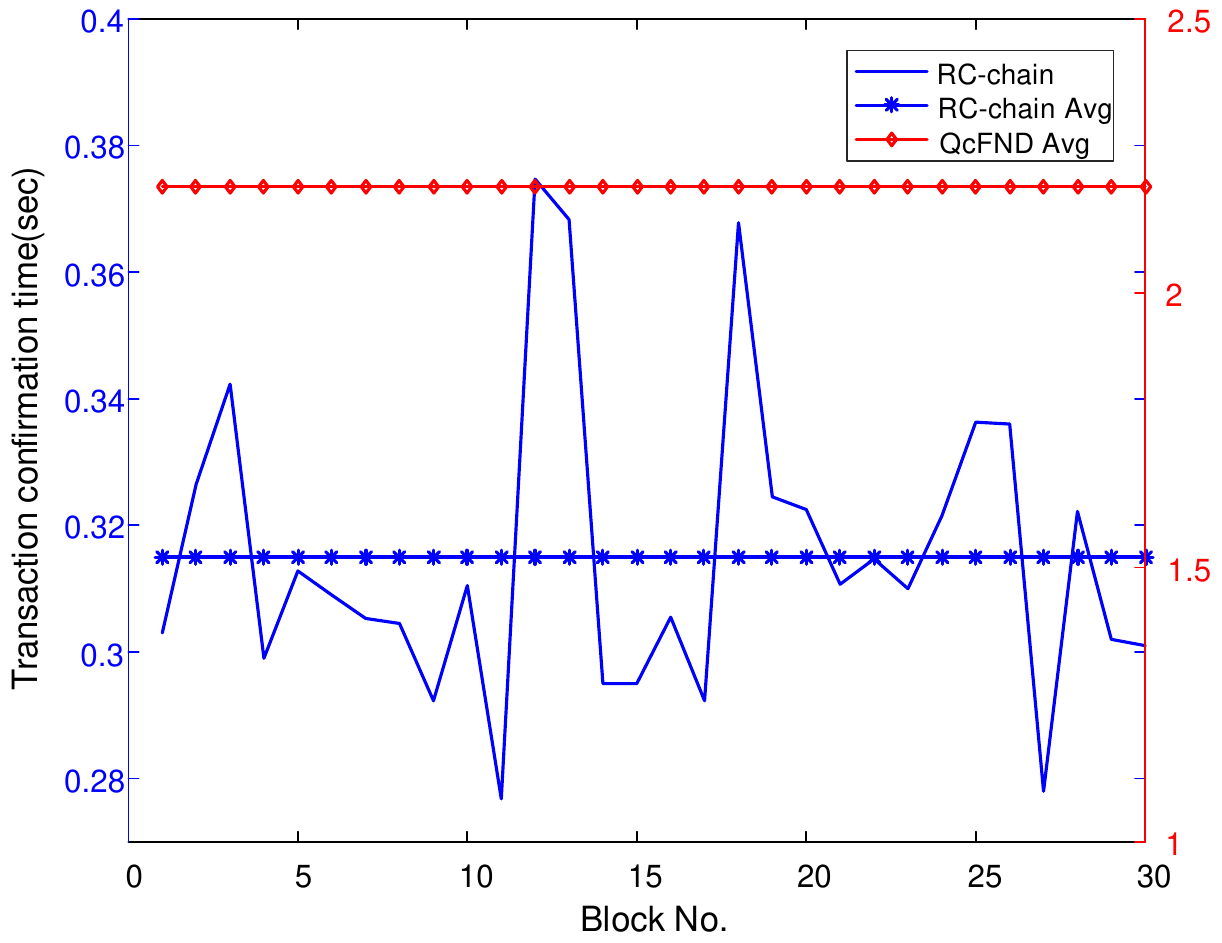}
	\caption{Transaction confirmation time of RC-chain and\\ QcFND.}
	\label{fig12}
\end{figure}

\par As mentioned in Section ~\ref{sec4.2}, the public blockchain restricts the transaction throughput due to its consensus mechanism (and other factors), which makes the transaction confirmation time longer. The block generation time for Bitcoin is 600s and the throughput is (1048576/500)/600 $ \approx $ 3.5 Tx/s. The average block time of Ethereum is 14.096s and the throughput is 71.428/14.096 $ \approx $ 6.6 Tx/s, the number of their miners are 2200 and 100 respectively \citep{memon2019simulation}. Most IoV services require low latency, and the throughput of RC-chain with 6 endorsing peers is 32 Tx/s – in other words, our results suggest that the RC-chain based consortium blockchain has distinct advantages compared other public blockchain schemes.

\section{Conclusion}
 \label{sec8}
 This paper proposed a novel RC-chain designed for constructing reliable VCS platform. We find that RC-chain efficiently leverages the resources of vehicular networks, provides transactions security, and enhances QoS over existing schemes. RC-chain has no third-party credit intermediaries; it realizes intelligent automatic transactions and stores trusted distributed ledgers according to predetermined rules based on smart contract. More importantly, RC-chain implements interconnection between organizations with shared interests, and achieves the transmission and accumulation of trust among VCS participants.
 
 We conducted a comprehensive set of experiments to find that RC-chain can be effectively applied, allows for swift transaction processing, and increases the throughput with lower execution cost compared to other schemes while satisfying trust constraints. Our numerical analysis also shows that the optimized TPFS model has favorable reputation updating effects over TWSL.

\begin{flushleft}
	\textbf{Acknowledgments}
\end{flushleft}
\par This research was supported by Opening Fund of Shandong Provincial Key Laboratory of Network based Intelligent Computing, National Natural Science Foundation of China under Grants No. 61802217, Natural Science Foundation of Shandong Province under Grants No. ZR201910280170, Project of Independent Cultivated Innovation Team of Jinan City under Grant No. 2018GXRC002, Project of Shandong Province Higher Educational Youth Innovation Science and Technology Program under Grant No. 2019KJN028.

\bibliographystyle{elsarticle-harv} 
\bibliography{mybib.bib}

\end{document}